\documentstyle[twocolumn,prl,aps,epsfig]{revtex}

\begin{document}
\draft
\twocolumn[\hsize\textwidth\columnwidth\hsize\csname @twocolumnfalse\endcsname
%
%
%

\title{Biquadratic interactions and spin-Peierls 
transition in the spin 1 chain LiVGe$_2$O$_6$}

\author{P. Millet $^1$, F. Mila $^2$, F. C. Zhang$^3$, M. Mambrini$^2$, 
A. B. Van Oosten$^2$, V. A. Pashchenko$^4$, A. Sulpice$^5$, A. Stepanov$^6$}
\address{$^1$ Centre d'Elaboration des Mat\'eriaux 
et d'Etudes Structurales, 29,
rue J. Marvig, 31055 Toulouse Cedex, France  \\
$^2$ Laboratoire de Physique Quantique, Universit\'e Paul Sabatier, 
31062 Toulouse Cedex, France\\
$^3$ Department of Physics, University of Cincinnati, OH 45221-0011\\
$^4$ High Magnetic Field Laboratory, Max-Planck Institut f\"ur
Fesk\"orperforschung/Centre National de la Recherche Scientifique, BP 166,
38042 Grenoble Cedex 9, France\\
$^5$ Centre de Recherches sur les Tr\`es Basses Temp\'eratures, 25 Avenue des
Martyrs, BP166, 38042 Grenoble Cedex 9, France\\
$^6$ Laboratoire Mat\'eriaux: Organisation et Propri\'et\'es, Universit\'e
d'Aix-Marseille III, Facult\'e des Sciences de St J\'er\^ome, Case 151, 13397
Marseille Cedex 20}

\date{\today}
\maketitle

\begin{abstract}
The magnetic susceptibility of a new one-dimensional, S=1 system, the vanadium 
oxide LiVGe$_2$O$_6$, has been measured. Contrary to previous S=1 chains, it
exhibits an abrupt drop at 22 K typical of a spin-Peierls 
transition, and it is 
consistent with a gapless spectrum above this temperature.
We propose that this behaviour is due to the presence of
a significant biquadratic exchange interaction, a suggestion supported by
quantum chemistry calculations that take into account the quasi-degeneracy of
the t$_{2g}$ levels.

\end{abstract}

\vskip2pc]
\narrowtext

The physics of one-dimensional (1D), spin 1 chains has attracted a considerable
amount of attention after the prediction by Haldane that 
the 1D Heisenberg model
has a spin gap for integer spins\cite{haldane}, a prediction confirmed since 
then by the 
observation of a gap in many spin 1 chains\cite{renard}. A lot of progress has 
been made by studying the most general Hamiltonian describing an isotropic 
coupling between neighbouring spins 1, namely\cite{affleck}
\begin{equation}
H=  J' \sum_i  \vec S_i.\vec S_{i+1} + J'' \sum_i (\vec S_i.\vec S_{i+1})^2
\label{hamiltonian}
\end{equation} 
The phase diagram of this model is extremely rich\cite{schollwock}. 
It is most easily 
described using the parametrization 
$J'=J \cos \theta, J''=J \sin \theta$. The system has two gapped phases: The
Haldane phase for $-\pi/4<\theta<\pi/4$ and a dimerized, gapped phase for
$-3\pi/4<\theta<-\pi/4$. The two phases are connected by a critical point at
$\theta=-\pi/4$ for which the model is solvable by Bethe ansatz and 
the spectrum gapless. In the following, we will use the other standard
parametrization $J'=J, J''=-\beta J$ because $\beta$ is a direct measure of the
relative strength of the biquadratic coupling to the bilinear one. With this
parametrization, the Haldane phase is limited to $J>0$ and $-1<\beta<1$, while
the dimerized, gapped phase corresponds to $\beta>1$ if $J>0$ and $\beta<-1$ if
$J<0$.
Hopelessly, it has not been possible so far to explore this phase diagram
experimentally due to the lack of systems with a sizable biquadratic exchange:
A well admitted point of view is that $\beta$ is always very small, and the 
quantitative interpretation of the experimental results obtained on many 
spin 1 chain compounds with the pure Heisenberg model has largely confirmed
this assumption.

In this Letter, we study the magnetic properties of a compound 
which has forced 
us to abandon this point of view, namely the vanadium oxide LiVGe$_2$O$_6$.
Although this system is clearly a spin 1 chain given its structure, its
susceptibility is markedly different from that of other spin 1 chains, 
and, as we shall explain below,
the most plausible explanation seems to be that there is a significant
biquadratic exchange.

\begin{figure}[hp]
\centerline{\psfig{figure=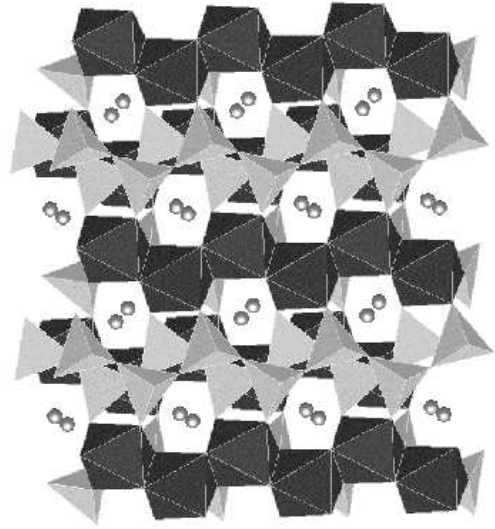,width=8.0cm,angle=0}}
\vspace{0.5cm}
\caption{Polyhedra representation of LiVGe$_2$O$_6$ structure down the a$^\ast$
  axis.}
\label{fig1}
\end{figure}

A powder sample of LiVGe$_2$O$_6$ was prepared by solid state reaction method
starting from a stoichiometric mixture of Li$_2$CO$_3$, GeO$_2$ and V$_2$O$_3$
which was placed in a platinum crucible and heated under vacuum 2 hours at
200$^\circ$ C and then 12 hours at 900$^\circ$ C. A second thermal treatment
at 800$^\circ$ C for 12 hours was necessary to obtain the pure compound.
Lithium vanadium metagermanate LiVGe$_2$O$_6$ is the end member of the solid
solution LiV(Si$_{2-x}$Ge$_x$)O$_6$ ($0\le x\le 2$) isolated 
and characterized by
Satto et al\cite{satto}. This compound belongs to the alkali metal pyroxene
family denoted AM$^{3+}$B$_2$O$_6$ (A=alkali, B=Si, Ge, M=a variety of cations
at a valence state $3+$) which has been extensively studied since 
many compounds
are naturally occuring minerals. The structure of LiVGe$_2$O$_6$ has been
determined from X-ray powder diffraction Rietveld analysis at 295 K. It
crystallises in the monoclinic system, space group P2$_1/$c compared to C2$/$c
for LiVSi$_2$O$_6$  with a=9.863(4)\AA, b=8.763(2)\AA, c=5.409(1)\AA and
$\beta$=108.24(1)$^\circ$. The structure, depicted in Fig. 1, is made of
infinite isolated chains of edge sharing VO$_6$ octahedra linked together by
chains of corner sharing GeO$_4$ tetrahedra. The consequence of the change in
the space group is that the chains of tetrahedra equivalent in LiVSi$_2$O$_6$
are distorted in LiVGe$_2$O$_6$ with two distinct 
crystallographic sites Ge1 and
Ge2. The geometric parameters are reported in Table 1. Note that the chains are
only connected to their neighbours through two GeO$_4$ tetrahedra so that the
coupling perpendicular to the chains must be very small.

\begin{figure}[hp]
\centerline{\psfig{figure=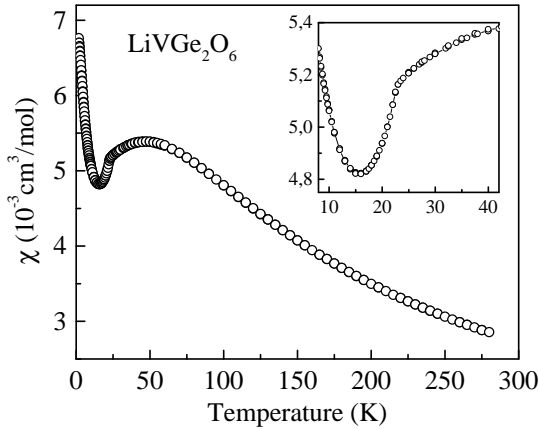,width=8.0cm,angle=0}}
\vspace{0.5cm}
\caption{Magnetic susceptibility of LiVGe$_2$O$_6$.}
\label{fig2}
\end{figure}

The magnetic susceptibility has been measured on a powder sample between 
1.75 K and 350 K and for magnetic fields ranging from 1 Oe to 7 T using 
a SQUID magnetometer. The results
are insensitive to the magnetic field strength, and  
the data are shown in Fig.2. 
Although the impurity contribution at low temperature is not negligible, two
features are clearly visible. 
The most spectacular one is an abrupt drop below 22 K. 
The behaviour below that temperature is typical of a spin-Peierls system
with impurities\cite{hase}, and it is very likely that this is indeed a
spin-Peierls transition. In particular, any explanation based on level crossings
induced by the magnetic field can be excluded since the results are independent
of the strength of the magnetic field. Besides, the susceptibility above this 
temperature is typical
of gapless systems like the Heisenberg S=1/2 chain, and not of spin 1 chains
with a Haldane gap, in which case the susceptibility drops much faster below
its maximum. In fact, since the spin-Peierls instability occurs
in gapless systems because of the magnetic energy to be gained by opening a
gap, both features are consistent with each other, and 
the main question is to understand the origin of the absence of a Haldane 
gap in the first place.

The first possibility is that although vanadium is in the oxydation state 3+, 
i.e. the electronic configuration is $3d^2$, one could have an effective spin
1/2 model, as was advocated a long time ago by 
Castellani et al\cite{castellani} in the context
of V$_2$O$_3$. To check this, 
we have fitted the high temperature data with the expression
$\chi(T)=\chi_{VV}+C/(T+\theta)$. A reasonable fit is obtained for
$\chi_{VV}=6.3\times10^{-4}$ cm$^3$/mol, $C=0,80$ cm$^3$K/mol 
and $\theta=79 K$. This value of $C$
is consistent with a spin 1 system with an average $g$ value of 1.79, a value
typical of spin 1 V$^{3+}$.
Besides, we have performed an ESR
investigation of this system. The only signal that we could 
detect is typical of
V$^{4+}$, but this signal corresponds to the small, paramagnetic contributions 
at low temperature. The fact that the main signal has not been detected is
consistent with a spin 1 state originating from V$^{3+}$ since any single-ion 
anisotropy pushes the resonance frequency outside the experimental range 
unless one works in a sufficiently high magnetic field. 
So it is clear that this
is indeed a spin 1 system.

\begin{figure}[hp]
\centerline{\psfig{figure=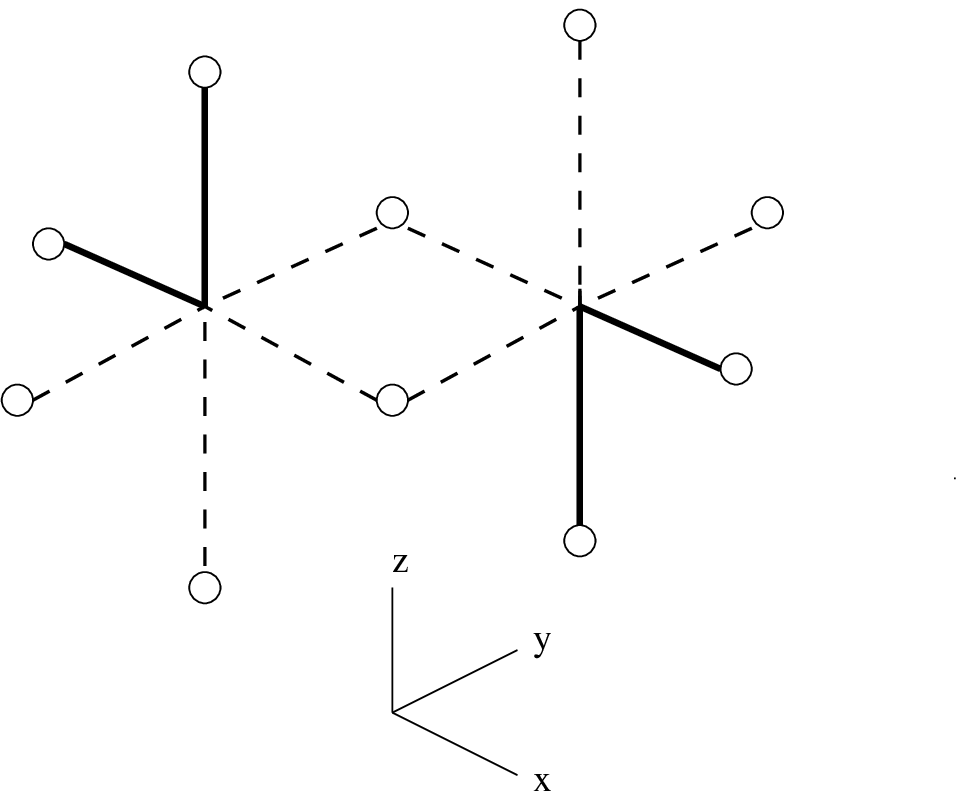,width=6.0cm,angle=0}}
\vspace{0.5cm}
\caption{Schematic representation of two-neighbouring V$^{3+}$ ions with their
surrounding oxygen octahedra. The solid (dashed) lines correspond to short
(long) bonds.}
\label{fig3}
\end{figure}

Taking for granted that the main contribution comes from spin 1 originating 
from V$^{3+}$, the possible explanations for the absence of a gap are:
i) The coupling between the chains; ii) A strong uniaxial, easy plane
anisotropy; iii) A significant biquadratic interaction. To review these
various possiblities, it is useful to start with a brief description of the
quantum chemistry of the vanadium ions.
In this compound, the vanadium
atoms are sitting in a distorted oxygen octahedron (see Fig. 3) with two
neighbouring VO bonds shorter than the other four. This situation is very
similar to the distortion of the sulfur octahedron surrounding vanadium in
AgVP$_2$S$_6$\cite{lee}, and the three $t_{2g}$ orbitals are split 
into a low-lying 
doublet
($d_{xy}$, $d_{yz}$) and a single orbital ($d_{xz}$) at an energy $\Delta$ 
above the doublet. Ab-initio calculations of the energy levels have been 
performed for a single ion embedded in a crystal, 
and the estimates for $\Delta$
are in the range 270-300 meV depending on the embedding. The degeneracy between
the $d_{xz}$ and $d_{yz}$ orbitals is actually lifted due to the low symmetry,
but this splitting is very small according to ab-initio calculations (in the
range 3 - 7 meV) and physically unimportant, so we will neglect it for
simplicity. The other on-site parameters are the Hund's rule coupling $J_H$
and the repulsion between electrons on the same (resp. different) orbitals 
$U_1$ (resp. $U_2$). Typical values are $J_H=0.5 - 1$ eV, $U_1=4 - 6$ eV and
$U_2=U_1-2 J_H$. Vanadium first neighours sit in edge-sharing octahedra, the
short bonds being opposite to each other and not 
involved in the common edge (see
Fig. 3). 
If we denote by $t_{ij}$ the hopping integrals between Wannier states
corresponding to the $t_{2g}$ orbitals with $i,j=1$, $2$ and $3$ for $d_{xy}$, 
$d_{yz}$
and $d_{xz}$ respectively, then $t_{12}=t_{13}=0$ by symmetry while $t_{11}$,
$t_{22}$ and $t_{33}$ involve direct overlaps between $3d$ orbitals and 
can be estimated from
standard tables\cite{harrison}: $t_{11}=0.3$ eV and $t_{22}=t_{33}=-0.1$ eV. 
Finally $t_{23}$ will
have a contribution from indirect hopping through the oxygen 2p orbitals and is
expected to be the largest hopping parameter in the problem, presumably in the
range $t_{23}=0.5 - 1$ eV. Let us now discuss the possible explanations for the
absence of a gap.

{\bf i) Interchain coupling:} 
It is well known that a relatively modest interchain coupling can close the 
gap\cite{sakai}. However, to have a significant effect 
on the susceptibility in 
the range 22-55K, the interchain coupling should be in that range.
Such a large ratio of the interchain
to intrachain coupling seems very unlikely given the very one-dimensional
structure of the compound. Besides if the interchain coupling was of the order
of 20 K or larger, one should find a transition to long range order in the same
temperature range, not a spin-Peierls transition. 
So we believe that the interchain coupling plays no significant role in the
present system.

{\bf ii) Uniaxial anisotropy:} Several theoretical studies have shown that the
presence of an easy plane, uniaxial anisotropy of the form $D\sum_i(S_i^z)^2$
can close the Haldane gap, the critical value being $D=J$ if the coupling
between the spins is purely Heisenberg and isotropic in spin 
space\cite{schulz}. Can $D$ be
of the order of 50 K in the present system? The answer is clearly negative 
on both experimental and theoretical grounds. Given the electronic structure of
V$^{3+}$ in the present compound, the spin-orbit coupling $\lambda \vec
L.\vec S$ gives rise to a uniaxial, easy-plane anisotropy of order
$\lambda^2/\Delta$ and a reduction of the perpendicular $g$ factor:
$g_\perp=2(1-\lambda/\Delta)$. In the isolated ion V$^{3+}$, 
$\lambda=104$ ${\rm cm}^{-1}$ , which
can be considered as an upper bound to its value in a crystal due to covalency
effects. Now the average $g$ factor determined from the high temperature
susceptibility is equal to 1.79. Assuming that $g_\parallel$ has the typical
value 1.9, this implies that $g_\perp=1.73$, i.e. $\lambda/\Delta=.13$. With
$\lambda=104$ ${\rm cm}^{-1}$, this leads to $D=20$ K, a value clearly 
too small to close
the gap. Besides, the actual value of $\lambda$, hence the value of $D$, is
certainly smaller than that. A careful analysis with 
single crystal NMR would be
very useful in getting more precise estimates. Theoretically, 
the large value of
$\Delta$ also points to a very small value of $D$, of the order of 
a few K. This
value is typical of V$^{3+}$ in this kind of environment\cite{takigawa}. So the
explanation in terms of uniaxial anisotropy can be discarded.

{\bf iii) Biquadratic interaction:} Thus, the only remaining possibility 
to explain 
the absence of a Haldane gap is the presence of a significant biquadratic 
interaction. 
Since the presence of a biquadratic interaction of significant magnitude  
has never been reported in any system,
the most important task is to see whether such an interaction 
could indeed occur
in the present case. So we have considered the general problem of the effective
coupling between two spins 1 in almost degenerate $t_{2g}$ orbitals. 
This turns out to
be a rather complicated problem with very different results in different
situations. So we will only present a summary of the results relevant for
LiVGe$_2$O$_6$. The details will appear in a forthcoming paper\cite{mila}.
Since the Hilbert space has a total of 495 states, this problem can be 
easily diagonalized numerically. As long as the energy $\Delta$ is not too
small, the low energy sector contains 9 states organized in 3 levels 
with degeneracy 1, 3 and 5,
typical of an isotropic coupling between spins 1. 
Now the splitting between these levels can be used to determine both the
bilinear and biquadratic exchange integrals since 
the energy levels of two spins
1 coupled by the Hamiltonian $J[\vec S_1.\vec S_2-\beta (\vec S_1.\vec S_2)^2]$
are given by $E_0=-J(2+4\beta)$, $E_1=-J(1+\beta)$ and $E_2=J(1-\beta)$ for
$S_{tot}=0, 1$ and $2$ respectively. It turns out that 
in a large parameter range,
and in particular for physically relevant parameters, 
$\beta$ can be larger than
1, i.e. the biquadratic interaction can be significant and even dominant. For
instance, if we take $\Delta=300$ meV, 
$U_1=5$ eV, $J_H=.5$ eV, $t_{11}=0.3$ eV 
and $t_{22}=t_{33}=-0.1$ eV, $\beta$ is already equal to .7 if $t_{23}=.5$ eV
and increases with $t_{23}$ until it diverges when
the bilinear exchange vanishes and 
becomes ferromagnetic.

This remarkable behaviour is best understood in the context of a
perturbation expansion in the hopping integrals. 
In the non-perturbed ground state each site carries two electons 
in a triplet state built out of the two low-lying orbitals. Then the 
degeneracy is in general lifted at second order, as for kinetic exchange, 
and the coupling is of the Heisenberg form, i.e. purely bilinear. What is 
important however is that it contains both ferromagnetic and antiferromagnetic 
contributions: $t_{11}$ and $t_{22}$ yield an AF contribution
$J_{AF}=(1/2)(t_{11}^2+t_{22}^2)(1/E_2+1/E_3)>0$, with $E_2=U_1+\Delta$ and
$E_3=U_1+2J_H+\Delta$ due to Pauli exclusion principle, while $t_{23}$ yields
a ferromagnetic contibution $J_{F}=(2/3)t_{23}^2(1/E_4-1/E_1)<0$ with
$E_1=U_2+\Delta$ and $E_4=U_2+2J_H+\Delta$ due to the
Hund's rule coupling that favours intermediate states with high spin. Then if
these couplings are of the same order, one has to go to fourth order to
calculate the exchange integrals. It turns out that at this order 
the coupling is no longer purely bilinear
but has a biquadratic contribution with a negative sign, and the ratio of this
contribution to the bilinear one diverges at the transition between
antiferromagnetic and ferromagnetic bilinear exchanges. The explicit 
expressions are very long and will be given in a forthcoming paper. In
general, $\beta$ becomes substantially large when the coupling from the second 
order contribution, $|J_{AF}+J_{F}|$, is comparable to $t^4/(U_1^2 \Delta)$, 
with $t$ 
the largest of the hopping integrals. The parameters we estimate for
LiVGe$_2$O$_6$ fall in this range.
So, whereas all other explanations could be discarded 
on the basis of elementary
but robust arguments, it seems that the presence of a significant biquadratic
interaction is a natural consequence of the relatively small splitting of the
$t_{2g}$ levels, i.e. of the quasi orbital degeneracy of the system. 

Note that the order of magnitude of the exchange integral supports by itself
this explanation: Without any competition between $J_F$ and $J_{AF}$ the
exchange integral would be much larger, like e.g. in AgVP$_2$S$_6$, where it 
is of the order of 400 K\cite{colombet}. 
As a consequence of this relatively small value of
the nearest neighbour coupling, the relative magnitude of the
next-nearest 
neighbour interaction, which has been implicitly neglected so far, 
might actually not be as small as what a naive guess based on the geometry
suggests. While such a term will, if anything, favour a dimerized groundstate, 
it does not close the Haldane gap\cite{kolezhuk} and cannot by itself account 
for the properties of the present system.

The presence of a biquadratic 
interaction has several interesting consequences.
First of all, it will modify the temperature dependence of the susceptibility,
and the data above the spin-Peierls transition are indeed consistent with a
large value of $\beta$. In particular, the susceptibility is quite flat below
the maximum, which is consistent with a very small gap, i.e. with a value of
$\beta$ around 1\cite{mambrini}. Besides, the gapped phase that occurs beyond 
$\beta=1$
corresponds to a two-fold, dimerized ground state, very much like for the 
spin 1/2 chain with large enough second neighbour exchange, and a Peierls
transition with a dimerization of the lattice is of course a very likely
instability of the system. So an explanation in terms of a biquadratic
interaction of the same order but larger than 
the bilinear one is at the present
stage the simplest and most natural explanation of the very peculiar properties
of LiVGe$_2$O$_6$. Further experimental investigations of this material to
confirm the Peierls nature of the transition and to study its magnetic
properties with and without magnetic field is likely to be an active field of 
research in the next few years.

We acknowledge useful discussions with P. Carretta, O. Golinelli, T. Jolicoeur,
E. Sorensen and F. Tedoldi. F. M. acknowledges the hospitality of the ETH
Z\"urich, where this project was completed.

\begin{table} [h]
\begin{center} 
\begin{tabular}{rlrlrll}
V-O5 & 1.89(1) & Ge1-O5 & 1.73(1) & Li-O1 & 2.04(1)\\
O3 & 1.94(1) &     O4 & 1.74(1)x2 & O3 & 2.08(1)\\
O1 & 2.04(1) & O1 & 1.78(1) & O4 & 2.11(1)\\
O1 & 2.06(1) & &  & O2 & 2.18(1)\\
O2 & 2.07(1) & Ge2-O6 & 1.75(1x2) & O5 & 2.33(1)\\
O2 & 2.08(1) & O3 & 1.76(1) & O6 & 2.49(2)\\
   &         & O2 & 1.77(1) & V-V & 3.149(3)\\
\end{tabular}
\vskip.5cm
\caption{Selected geometrical parameters (\AA)}
\end{center}
\end{table}

\end{document}